\documentclass{PoS}

\title{ QED with heavy ions: on the way from strong to supercritical fields  }

\ShortTitle{QED with heavy ions: on the way from strong to supercritical fields}

\author{\speaker{V.M.~Shabaev}, A.I.~Bondarev, D.A.~Glazov, Y.S.~Kozhedub, I.A.~Maltsev, 
  A.V.~Malyshev, R.V.~Popov, D.A.~Tumakov, and I.I.~Tupitsyn  \\
        St. Petersburg State University\\
        E-mail: \email{v.shabaev@spbu.ru}}

\abstract{The current status of tests of quantum electrodynamics with heavy ions is reviewed.
  The theoretical predictions for the Lamb shift and the hyperfine splitting in heavy ions are compared with
  available experimental data. Recent achievements and future prospects in studies of the $g$ factor with
  highly charged ions are also reported. These studies can provide precise determination of the fundamental
  constants and tests of QED within and beyond the Furry picture at the strong-coupling regime. Theoretical calculations
  of the electron-positron pair creation probabilities in low-energy heavy-ion collisions are also considered. Special attention
  is paid to tests of QED in supercritical-field regime, which can be accessed
  in slow collisions of two bare nuclei with the
  total charge number larger than the critical value, $Z_{\rm crit} \approx 173$. In the supercritical field,
  the initially neutral vacuum can spontaneously decay into the charged vacuum and two positrons.
  It is demonstrated that  this fundamental phenomenon
  can be observed via impact-sensitive measurements of the pair-production probabilities.}

\FullConference{International Conference on Precision Physics and Fundamental Physical Constants - FFK2019\\
		9-14 June, 2019\\
		Tihany, Hungary}

\begin{document}

\section{Introduction}

Bound-state quantum electrodynamics (QED) has been tested for many decades with light atomic systems,
such as hydrogen, helium, positronium, and muonium. As the result of these tests, nowadays nobody doubts
the basic principles of QED,  including the quantization of the interacting electron-positron and electromagnetic
fields as well as the renormalization procedure. However, this does not  mean that any further QED tests are not needed.
First, the studies of light atomic systems restrict tests of the QED methods to few low orders in the parameters
$\alpha$ and $\alpha Z$, where $\alpha$ is the fine structure constant and $Z$ is the nuclear charge number.
Even if we believe that these methods should also work to higher orders, this
can not be fully guaranteed unless it is proven by comparison of experiment and theory. This means that any further
advance in precision of both theory and experiment is of great importance. Moreover, high-precision measurements
with heavy few-electron ions, which became possible in the last few decades, provide a unique
opportunity to significantly expand the region of QED tests. Namely, the study of these systems allows us
to test  bound-state QED in the nonperturbative in  $\alpha Z$ regime (in other words,
to all orders in  $\alpha Z$). Again, although it is expected that the standard QED formalism can be naturally
adapted for calculations of these systems, the corresponding tests in the nonperturbative regime
must be performed before  these studies can be used for various practical applications.
As the main reference points for the QED tests in the nonperturbative in $\alpha Z$ regime, one may consider
the Lamb shift, the  hyperfine splitting, and the $g$ factor of heavy few-electron ions. The current status
of these tests is briefly reviewed in the first part of the present paper.

The methods developed for precise QED calculations of heavy few-electron ions can also serve as an important
intermediate step towards tests of QED in supercritical Coulomb field. Such  field can be created
in low-energy collisions of two heavy ions with the total nuclear charge number
exceeding the critical value, $Z_1+Z_2>Z_{\rm crit} \approx 173$.
In the supercritical field the lowest one-electron quasimolecular
 level enters into the negative-energy continuum
and becomes a resonance.  If this level is empty, its ``diving'' into the negative-energy continuum
should result in the decay
of the originally neutral vacuum into the charged vacuum and two positrons.
In the second part of the paper, we demonstrate that this fundamental
phenomenon can be observed in studying the impact-sensitive pair-production probabilities.

\section{Strong field QED with heavy ions}

\subsection{Lamb shift}
\begin{figure}
\begin{center}
  \includegraphics[width=0.2\textwidth]{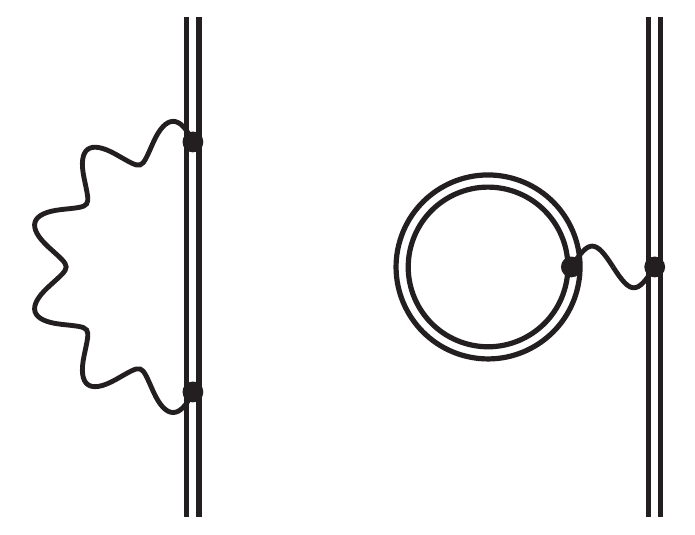}
\end{center}
  \caption{One-loop QED diagrams.}
\label{fig:1}       
\end{figure}

\begin{figure}
\begin{center}
  \includegraphics[width=0.5\textwidth]{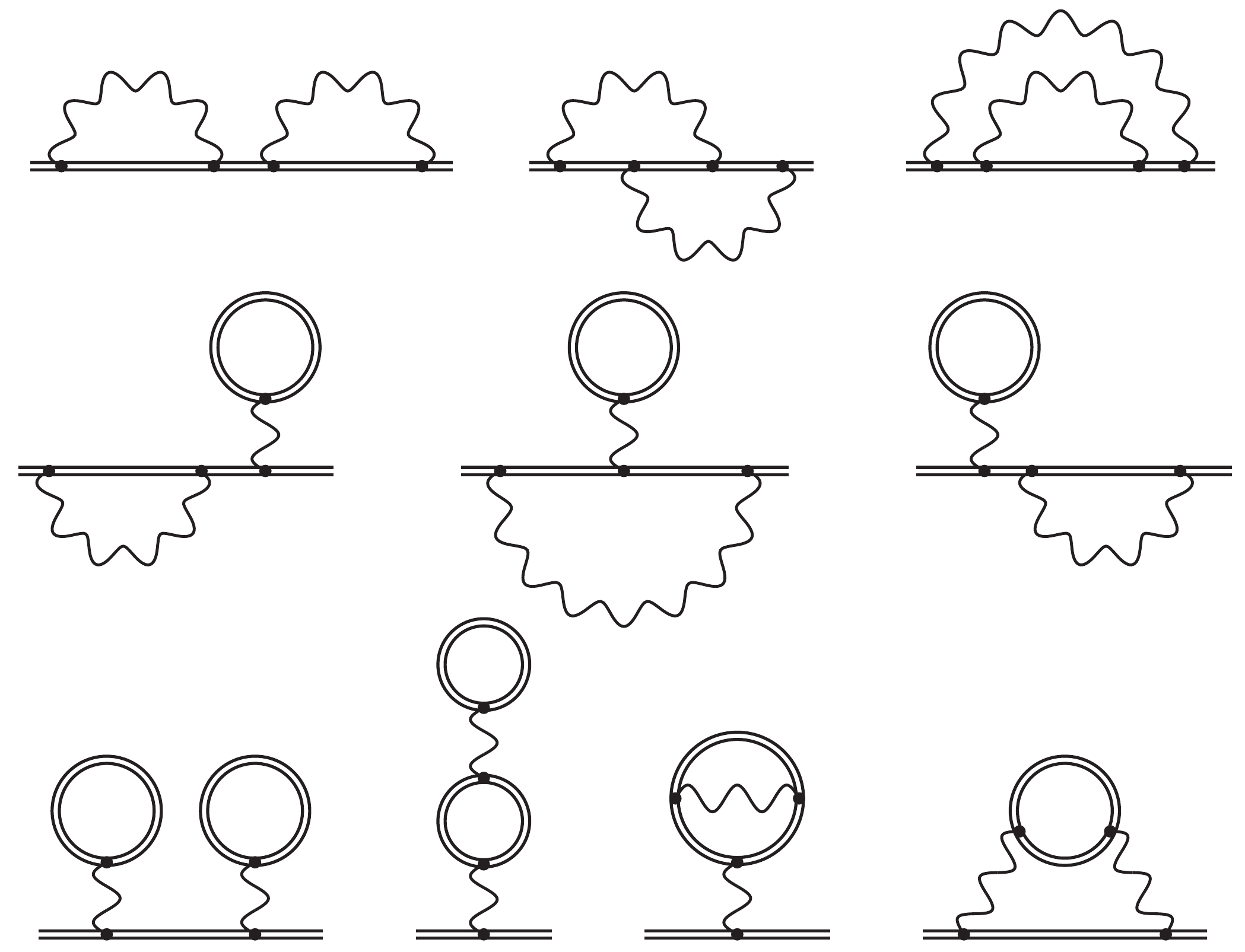}
\end{center}
  \caption{Two-loop QED diagrams.}
\label{fig:2}       
\end{figure}

The Lamb shift in H-like ions is mainly determined by the QED contributions
evaluated within the Furry picture. In this picture, the nucleus is considered
as a source of the external Coulomb field and the standard QED formalism
in the presence of  static classical field can be used. 
The main QED contributions to the Lamb shift are defined by one-loop
self-energy and vacuum-polarization diagrams which are presented in
 Fig.~\ref{fig:1}.
In contrast to light atoms, where the parameter $\alpha Z$ is small,
in highly charged ions these diagrams must be calculated without any expansion
in $\alpha Z$. Nowadays these calculations cause no problem.
Much more challenging is the calculation of  two-loop QED corrections which are defined
by the diagrams depicted in Fig.~\ref{fig:2}.
To date, most of these diagrams have been
calculated and the total uncertainty of the two-loop contribution is determined
by the last two diagrams  in Fig.~\ref{fig:2},
which have been evaluated in the free-electron-loop
approximation only. In addition to the Furry-picture QED  contributions, one
has to take into account the nuclear recoil, nuclear size, and nuclear polarization
effects. The evaluation of the  nuclear recoil effect requires using
the QED formalism in the  nonperturbative
in $\alpha Z$ regime beyond the Furry picture. On the one hand, this
provides us a rather serious conceptual problem, but, on the other
hand, it gives a unique opportunity to test QED in a new region:
strong-coupling regime beyond the Furry picture. The nuclear
size effect can be easily evaluated by solving numerically the Dirac
equation with an extended nucleus. Finally, one should take into account
the nuclear polarization effect which can be described by two-photon-exchange
electron-nucleus interaction diagrams, in which intermediate nuclear states are excited.
The references to the calculations of all aforementioned contributions
can be found, e.g., in Ref. \cite{yer15}. In Table~\ref{tab:1}, we present the individual contributions
to the ground-state Lamb shift in H-like uranium, which should be considered
as the main reference point for the QED tests at strong fields.
Here the Lamb shift is defined
as the difference between the total binding energy
and the Dirac binding energy calculated for the pure Coulomb field
of the infinite mass nucleus, $V_{\rm C}=-\alpha Z/r$ . It should be noted
that in Ref.  \cite{yer15} the leading recoil correction was not
included in the Lamb shift to make the definition suitable
for both light and heavy ions.
The nuclear charge radius was taken from Refs. \cite{koz08,ang13}.
The finite nuclear size contribution was evaluated
taking into account the nuclear deformation effect \cite{koz08}.
It can be seen that the present status of theory and experiment
on the Lamb shift in H-like uranium
provides a test of QED in the nonperturbative regime at the 2\% level.

\begin{table}
\caption{Theoretical contributions to the $1s$ Lamb shift in H-like $^{238}$U, in eV.}
\label{tab:1}       
\begin{center}
\begin{tabular}{ll}\hline
Contribution  & Value \\ \hline
Furry picture QED  & 265.07(48) \\
Finite nuclear size  &  198.51(19) \\
Nuclear recoil   &  $\;\;\;\,\;$0.46(1)  \\
Nuclear polarization   & $\,\,-$0.20(10)  \\
Total theory   & 463.84(53)  \\
Experiment \cite{gum05}  & 460.2(4.6)  \\ \hline
\end{tabular}

\end{center}

\end{table}

A higher experimental accuracy was achieved for heavy
lithiumlike ions \cite{sch91,bra03,bei05}. The comparison of theory and experiment on the
$2p_{1/2} - 2s$ transition energy in Li-like uranium provides a test of QED
at the 0.2\% level (see Ref. \cite{koz08} and references therein).

\subsection{Hyperfine splitting}

The hyperfine splitting (HFS) of atomic levels is caused by the interaction of the atomic
electrons with the magnetic field induced by a nonzero nuclear magnetic moment.
For heavy few-electron ions this field becomes extremely strong.  For instance, in the case
of H-like $^{209}$Bi, the electron  experiences on
average the magnetic field of about 30 000 T.
This value is three orders of magnitude greater than that induced by the strongest superconducting magnets.
 It means that the study of the HFS with heavy few-electron ions
could provide tests of QED in a unique combination of the strongest electric
and magnetic fields. This has triggered a great interest to measurements of the HFS
in H-like ions \cite{exp1,exp2,exp3,exp4,exp5}. However, the calculations of the HFS
in H-like ions \cite{sha97}
revealed that the uncertainty due to the nuclear magnetization distribution
correction (so-called Bohr-Weisskopf effect) is generally of the same order of
magnitude as the QED correction
and, therefore, tests of QED effects on the HFS by the direct comparison of theory
and experiment for H-like ions are not possible. To circumvent this problem, in Ref.
\cite{sha01} it was proposed to study a specific difference of the HFS values
in H- and Li-like ions of the same heavy isotope,
\begin{eqnarray}
\Delta'E=\Delta E^{(2s)}-\xi \Delta E^{(1s)},
\end{eqnarray}
where  $\Delta E^{(1s)}$ is the HFS in the H-like ion,
$\Delta E^{(2s)}$ is the HFS in the corresponding Li-like ion, and
$\xi$ is chosen to cancel the Bohr-Weisskopf effect.
In Ref. \cite{sha01} it was shown that both the
parameter  $\xi$ and the specific difference $\Delta'E$
are very stable with respect to possible variations of
the nuclear models and nuclear parameters. This means
that both  $\xi$ and $\Delta'E$ can be evaluated to a high accuracy.
For instance,  in the case of  $^{209}$Bi the precise calculation
leads to $\xi$ = 0.16886.
 This method has a potential to test QED on the level of a few percent, provided 
the HFS is measured to accuracy $\sim 10^{-6}$.
This proposal has initiated precise experiments on the HFS in
Li-like  bismuth. As a result, in Ref. \cite{ull17} the experimental
value for the specific difference in  $^{209}$Bi was reported,
but the result was 7$\sigma$ off from the latest theoretical prediction
\cite{vol12}. This discrepancy established a
``hyperfine puzzle'' \cite{kar17},  which  was resolved in
Ref. \cite{skr18}. New calculations  of  the magnetic shielding factor in
$^{209}$Bi(NO$_3$)$_3$ and in $^{209}$BiF$_6^-$ performed in
Ref. \cite{skr18}
clearly showed that 
the nuclear magnetic moment of  $^{209}$Bi  widely used in literature is incorrect.
The new calculations and measurements for $^{209}$BiF$_6^-$  \cite{skr18}  lead to a new value
of the nuclear magnetic moment:  $\mu/\mu_N=4.092(2)$.
The individual contributions to the specific HFS difference, obtained with
the new magnetic moment, are presented  in Table~\ref{tab:2}. It can be seen that
the theoretical result agrees well with the experiment. However, the present status
of the accuracy of the nuclear magnetic moment and the HFS experiment
limits the QED tests to the 15\% level. More precise measurements
are  needed to provide stringent QED tests. 

\begin{table}
\caption{Theoretical contributions to { $\Delta' E =\Delta E^{(2s)}-\xi \Delta E^{(1s)}$ }
  in $^{209}$Bi (in meV)
  for $\mu/\mu_N=4.092(2)$  \cite{skr18}.
  All theoretical values are obtained by scaling  the related
contributions from Ref. \cite{vol12}.}
\label{tab:2}
\begin{center}

  \begin{tabular}{ll}\hline
   Contribution  & Value \\ \hline 
Dirac value  &  $-$31.665 \\ 
Interelectronic interaction, $\sim 1/Z $ &  $-$29.859 \\ 
Interelectronic interaction, $\sim 1/Z^{2+}$ & $\,\,\,\,\,\,\,\,$  0.254(3) \\
One-electron QED & $\,\,\,\,\,\,\,\,$ 0.036 \\
Screened QED & $\,\,\,\,\,\,\,\,$  0.192(2) \\ 
Total theory &  $-$61.043(5)(30) \\
Experiment \cite{ull17} & $-$61.012(5)(21) \\ \hline
\end{tabular}

\end{center}

\end{table}

\subsection{Bound-electron $g$ factor}

First high-precision measurements of the $g$ factor of highly charged ions
were performed for H-like carbon two decades ago \cite{haf00}. This experiment has
triggered a great interest to precise QED calculations of the bound-electron
$g$ factor in highly charged ions (see, e.g., Refs. \cite{sha15,har18} for reviews).
The measurements of the bound-electron $g$ factors in low- and middle-$Z$
H-like ions and comparison with the corresponding theory lead to the most
precise determination of the electron mass \cite{stu14,zat17,cza18}.
Recent measurements for two isotopes of Li-like calcium
 \cite{koe16}  allowed the first test of the relativistic theory of the nuclear recoil effect
with highly charged ions in presence of  magnetic field \cite{sha17}. 
High-precision measurements of the g factors of heavy few-electron ions
are anticipated in the near future at the Max-Planck Institut
fuer Kernphysik (MPIK)  in Heidelberg and at the HITRAP/FAIR facilities
in Darmstadt. In addition to the strong-field QED tests within the Furry
picture,  these measurements, combined with the corresponding calculations,
can provide a test of the QED theory of
the nuclear recoil effect on a few-percent level \cite{mal17}.
This would mean the first test of bound-state QED at the strong-coupling regime
beyond the Furry picture.
Measurements of the g factor of ions with nonzero nuclear spin, which are planned
at the same facilities, will give an access to the nuclear $g$ factors.
Precise values of the nuclear magnetic moments, which can be determined from these
experiments by means of the related theory  \cite{sha15,mos08}, would be of great
importance for the HFS study discussed above.
Finally, one should mention a possibility for an independent determination of the fine structure
constant from the g-factor experiments with highly charged ions, provided the corresponding
theoretical calculations are performed to the required accuracy \cite{sha06,yer16}.

\section{QED in supercritical fields}

In accordance with the presently accepted QED formalism, strong static electric
field can create electron-positron pairs, provided its strength exceeds a critical value.
This effect was predicted many decades ago but has never been observed experimentally.
Starting with Refs. \cite{sau31,hei36,sch51}
predicted spontaneous creation of electron-positron
pairs by a strong uniform time-independent  electric field (so-called Schwinger
mechanism) and
Refs. \cite{ger69,pie69}
predicted a similar effect
in a supercritical Coulomb field, great
efforts were undertaken  to find feasible scenarios for experimental observation of this
fundamental phenomenon. In particular, it was expected that the desired field  strength
could be achieved with new laser technologies (see, e.g., Ref. \cite{ale18} and references
therein). However, even in the most optimistic scenarios for the developments of these technologies,
the field strength which can be achieved in the not too distant future
is by two orders of magnitude smaller than the value needed to observe the Schwinger effect.
In what follows, we consider the pair creation by  supercritical Coulomb field.

\begin{figure}
\begin{center}
  \includegraphics[width=0.5\textwidth]{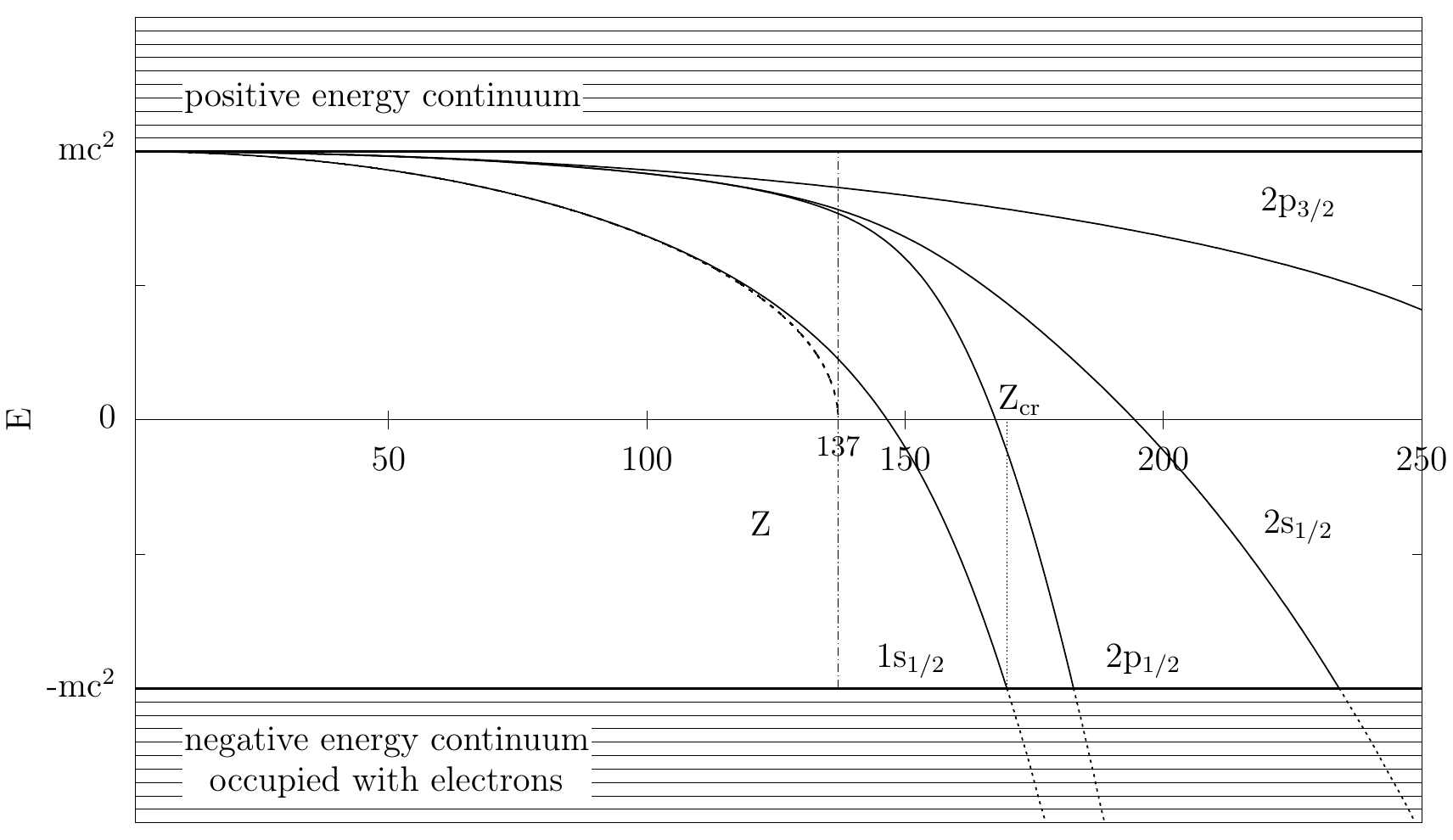}
\end{center}
  \caption{The low-lying energy levels of a H-like ion as functions of the nuclear charge number $Z$.}
\label{fig:3}       
\end{figure}

In  Fig.~\ref{fig:3}
we display the low-lying energy levels of a H-like ion as functions
of the nuclear charge number $Z$. In the case of the point-charge nucleus, the $1s$ level exists
up to $Z\approx 137$ and then disappears. For an extended nucleus, it goes continuously
down and reaches the negative-energy continuum at  $Z\approx 173$. If this level is
empty, its ``diving'' into the negative-energy continuum should lead to the decay of the
originally neutral vacuum into the charged vacuum and two positrons. Since there are no
nuclei with so high $Z$, the only way to access the supercritical regime is to study
low-energy collisions of two ions with the total nuclear charge larger than the
critical value, $Z_1+Z_2>Z_{\rm crit} \approx 173$. The corresponding experiments
were first performed about three decades ago at GSI (Darmstadt).
However, these experiments could not prove the spontaneous pair creation.
One of the reasons preventing the study of the effect of interest
consists in the fact that the experiments were performed with many-electron systems.
The investigations could be much more successful if they were performed for
collisions of bare  nuclei. New studies of low-energy heavy-ion collisions
at the supercritical regime are anticipated 
at the upcoming accelerator facilities in Germany, Russia, and China
\cite{gum09,ter15,ma17}. These facilities
will allow to perform  experiments on low-energy collisions of heavy bare nuclei.
However, there is another problem that poses a serious obstacle to observing the vacuum decay.
This  is a too short
period of time of the supercritical regime.
For instance, in U-U collisions at the energy
near the Coulomb barrier the supercritical regime emerges for about 10$^{-21}$s only. This
is by two orders of magnitude smaller than the time required for the spontaneous
pair creation. The world-leading Frankfurt's group, which worked on this topic for more than 20 years,  
finally concluded that the spontaneous pair creation
could only be observed in collisions with nuclear
sticking, in which the nuclei are bound to each other for some period of time by nuclear
forces \cite{gre85,rei05}. Since to date there is no evidence of existence of the
nuclear sticking in collisions of interest, this scenario does not seem promising.
One may expect, however, that the detailed study of quantum dynamics of the electron-positron
field  in low-energy heavy-ion collisions can result in finding some signatures
for the supercritical-field regime. To perform these studies, first of all
new methods, which would allow the calculations beyond the approximations
employed by the Frankfurt's group, were needed.
To this end, new efforts were
started a decade ago by the St. Petersburg's group with the developments
of new methods for the
relativistic calculations
of the charge-transfer, electron excitation and ionization processes 
\cite{tup10,tup12,dey12,dey13,mal13}. These methods can  be
easily adapted to calculations of the pair-creation probabilities
within the monopole approximation, which was mainly used 
by the Frankfurt's group \cite{mul88}. In this
approximation, the two-center nuclear potential is expanded in spherical
harmonics around the center of mass and then only
the zero-order spherical harmonic term is used in the calculations
(see, e.g., Ref. \cite{mal15} and references therein).
However, till recently it was not completely clear if the monopole
approximation is good enough to be used for studying in detail
the pair-creation processes in subcritical and supercritical regimes.
A significant progress in this direction was made recently in Refs.
\cite{mal17a,pop18,mal18}, where the effects beyond the monopole
approximation were evaluated by different methods.
These calculations revealed that, for the nuclear
charge numbers and the energies of interest,
this approximation works very well at small impact parameters. 
In particular,  the calculations at the energy near the Coulomb barrier showed that
the difference between the exact and the monopole-approximation results varies from
about 6\% at the zero impact parameter, $b = 0$, to about 10\% at $b = 10$ fm.
Thus, this approximation should work well for studying the energy and
impact-parameter dependences of the
pair-creation probabilities for the near-head-on collisions.

\begin{figure}
\begin{center}
  \includegraphics[width=0.5\textwidth]{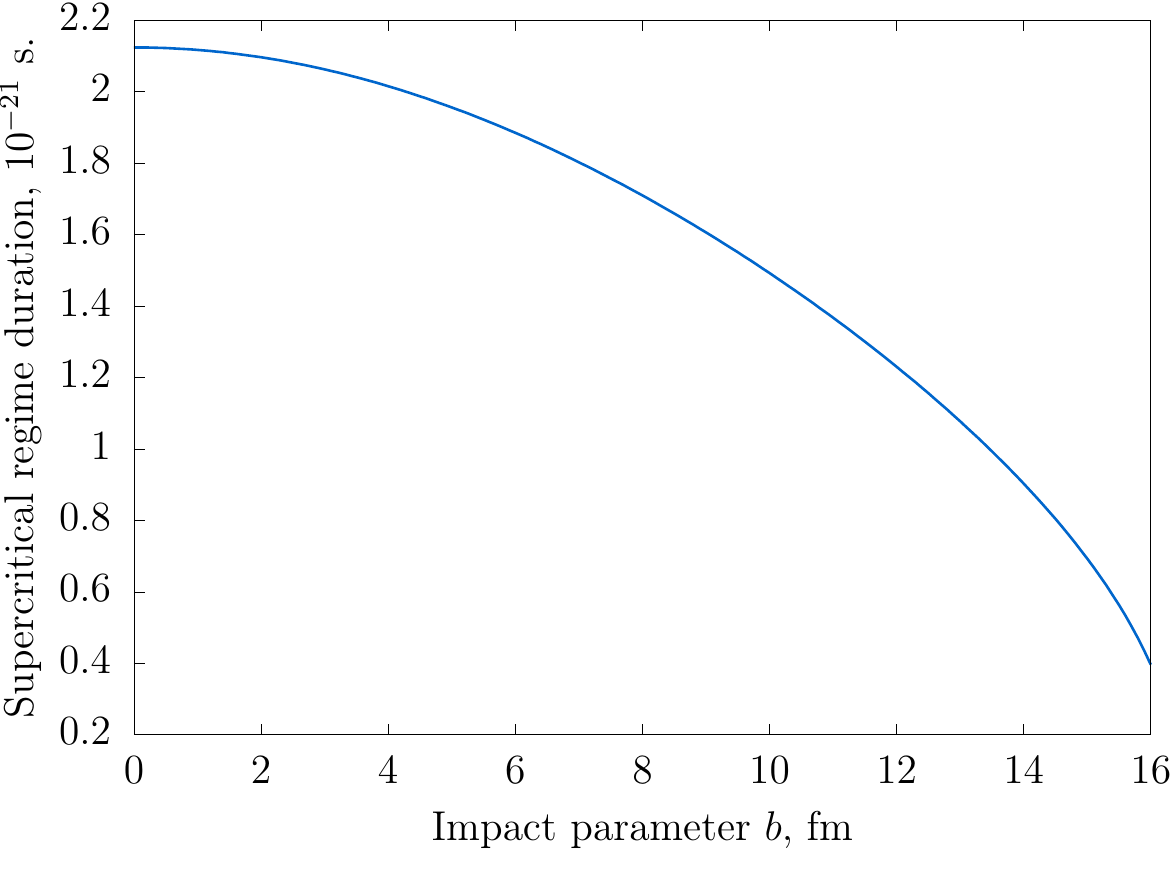}
\end{center}
\caption{The time period of the supercritical regime in collisions of bare uranium nuclei
  as a function of the impact parameter $b$ for a given
  minimal distance of the nuclear approach for all $b$,  $R_{\rm min}=16.5$ fm.}
\label{fig:4}       
\end{figure}

\begin{figure}
\begin{center}
  \includegraphics[width=0.5\textwidth]{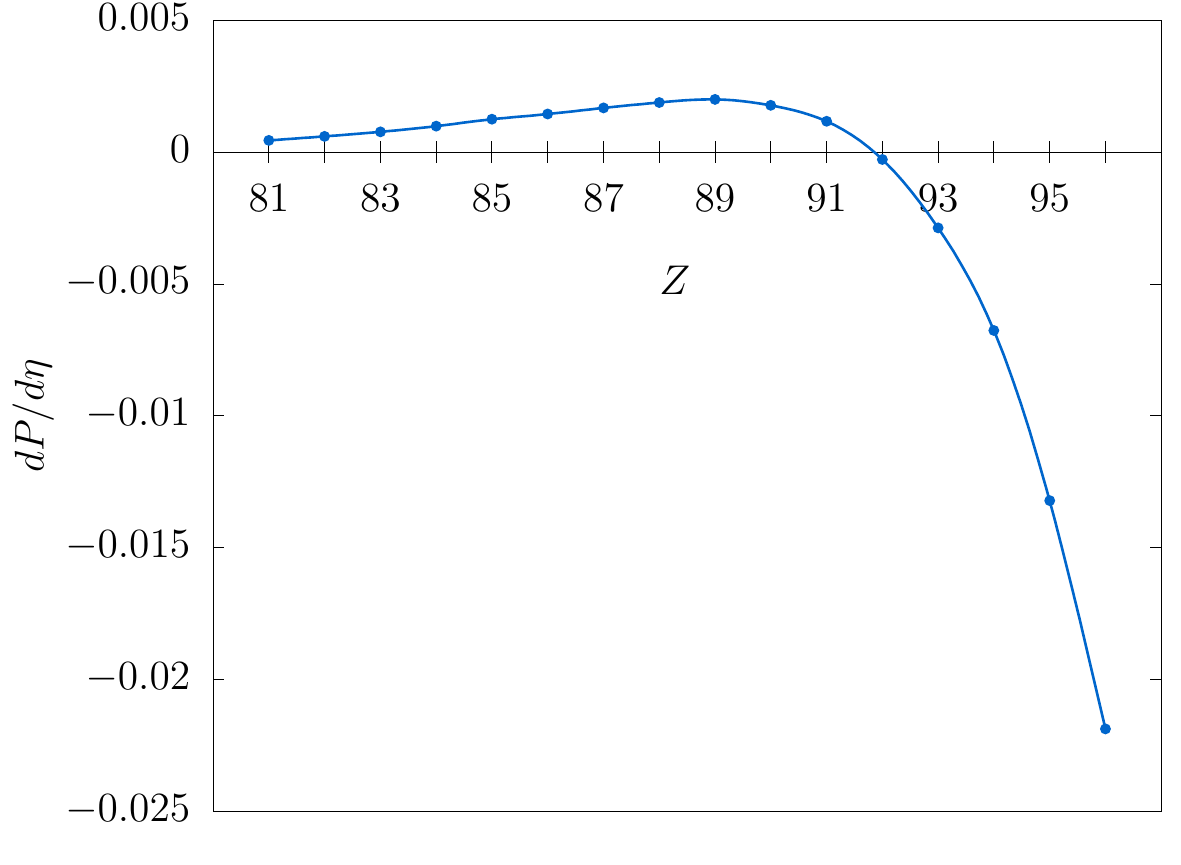}
\end{center}
\caption{ The derivative of the pair-production probability 
 $dP(\eta)/d\eta|_{\eta=1}$}  as a function of the nuclear
  charge number $Z$  for symmetric collisions,  $Z=Z_1=Z_2$.
  Here  $\eta=E/E_0$, $E$ is the collision energy, and $E_0$ is the head-on
  collision energy. All the values correspond to a given minimal distance of the nuclear
  approach, $R_{\rm min}=16.5$ fm (see the text).
\label{fig:5}       
\end{figure}

Let us now discuss how the impact-sensitive measurements of the pair-production probabilities
can provide observation of the qualitative difference between the subcritical and
supercritical regimes \cite{mal19}.
We consider a low-energy collision of two heavy nuclei with nuclear
charge numbers $Z_1$ and  $Z_2$ and the total charge number  larger than
the critical one, $Z_1+Z_2>173$.
To separate the spontaneous pair creation from the dynamical (induced) one,
let us consider the trajectories which correspond to the same
minimal distance between the nuclei ($R_{\rm min}$) but different impact parameters ($b$).
According to the Rutherford kinematics, these parameters are related by
\begin{eqnarray} \label{b}
b^2=R_{\rm min}^2 - R_{\rm min}\frac{Z_1 Z_2}{E}\,,
\end{eqnarray}  
where $E$ is the collision energy. At given value of $R_{\rm min}$, the minimal energy
corresponds to the head-on collision and
reads as
\begin{eqnarray}
E_0=\frac{Z_1 Z_2}{R_{\rm min}}\,.
\end{eqnarray}
Let us consider the U-U collision, for which the supercritical regime emerges when the nuclei approach
each other at the distance closer than 32.6 fm. With $R_{\rm min}=16.5$ fm, which corresponds to
the distance at which the nuclei are in 1-2 fm away from touching each other, the time period of
the supercritical regime as a function of the impact parameter $b$ is  presented 
in Fig.~\ref{fig:4}.
We note that, according to Eq. (\ref{b}), for a given value of  $R_{\rm min}$
the impact parameter $b$ and the energy $E$ are related to each other. As it can be
seen from Fig.~\ref{fig:4},  the more is $b$, the less is the supercritical-regime time period.
This corresponds to increasing the energies and, therefore, the relative velocity of the
colliding nuclei with increasing $b$. It is evident that the dynamical pair creation
must monotonically decrease with decreasing the velocity (and, therefore, $b$), provided
$R_{\rm min}$ is fixed. At the same time, the spontaneous pair creation must monotonically
increase with decreasing $b$, since the supercritical-regime time period increases.
This means that any increase of the pair-creation probability as a function of $\eta = E/E_0$
at  $\eta\rightarrow 1$ would reveal the effect of the spontaneous pair creation, which
is possible only in the supercritical regime. Moreover, even a qualitative change in
the behaviour of the derivative of the pair-production probability $P(\eta)$ with respect
to $\eta$ when going from the subcritical to supercritical regime would demonstrate the effect
of the spontaneous pair creation. In  Fig.~\ref{fig:5}  we present the behaviour of $dP(\eta)/d\eta|_{\eta=1}$
as a function of the nuclear charge number $Z$ for symmetric collisions  ($Z=Z_1=Z_2$).
It can be seen that even in the case of the U-U collision the effect of the decrease of
the derivative compared to the subcritical case ($Z<88$) is very pronounced.
We note also that an estimate of the bound quasimolecular level occupation probability
for the collision of a bare uranium nucleus with
a neutral uranium atom,  based on the methods developed in Refs. \cite{tup10,tup12,koz14},
shows that the filled K shell of the atom  can only lead to a few times decrease
of the pair-production probability, leaving the main conclusions qualitatively unchanged.
This makes feasible the experimental study of the pronounced
decrease of  $dP(\eta)/d\eta|_{\eta=1}$ as a function of $Z$ with
the near future facilities \cite{les16,hag19}. The observation of 
this decrease must be considered as a clear proof of the fundamental phenomenon:
the vacuum decay in supercritical Coulomb field.

\section{Conclusion}
We have reviewed the present status of tests of QED with heavy ions. The high-precision
measurements of the Lamb shifts in heavy H- and Li-like ions serve as the main reference
points of the QED tests at strong Coulomb field. The high-precision
HFS and $g$-factor experiments with heavy few-electron ions, which are anticipated
in the near future, should provide the QED tests in a unique combination of the
strongest electric and magnetic fields as well as tests of QED at the strong-coupling
regime beyond the Furry picture. Given these tests are performed, the experiment and
theory with heavy ions can be employed for precise determinations of various
nuclear parameters and fundamental constants.
The study of the  pair-creation probabilities in low-energy heavy-ion collisions
near the Coulomb barrier energy should provide a unique access to QED in supercritical
field. The precise measurements of the pair-creation probabilities for different impact
parameters but with a given minimal distance of approach of the colliding nuclei and comparison
with the theory
will result  either in discovery of the vacuum decay in supercritical Coulomb field
or in finding new physics, which is beyond the standard QED formalism.

\section*{Acknowledgements}
This work was supported by the Russian Science Foundation (Grant No. 17-12-01097).

\end{document}